\documentclass[12pt]{article}
\usepackage{amsmath,,calrsfs}
\usepackage{amsfonts}
\usepackage{amssymb}
\usepackage{amscd}
\usepackage{bbm}
\usepackage{fancybox}
\usepackage{cite}
\usepackage{amsmath,amsfonts,amsbsy}
\usepackage{pstricks,pst-node}
\usepackage[small,bf,hang]{caption2}
\usepackage{graphicx}
\usepackage{epsfig}
\usepackage{psfrag}
\usepackage{comment}

\usepackage{float}

\psset{unit=1.3cm,linewidth=.5pt,radius=.2}  

\usepackage{multirow}                     
\usepackage{float}                          
\usepackage{lscape}                         
\usepackage{bm}

\newcommand{\beq}{\begin{equation}}
\newcommand{\eeq}{\end{equation}}
\newcommand{\bi}{\begin{itemize}}
\newcommand{\ei}{\end{itemize}}
\newcommand{\bt}{\begin{tabular}}
\newcommand{\et}{\end{tabular}}
\newcommand{\bc}{\begin{center}}
\newcommand{\ec}{\end{center}}

\newcommand{\be}{\begin{equation}}
\newcommand{\ee}{\end{equation}}
\newcommand{\bea}{\begin{eqnarray}}
\newcommand{\eea}{\end{eqnarray}}
\newcommand{\ba}{\begin{array}}
\newcommand{\ea}{\end{array}}

\def\bbox{{\,\lower0.9pt\vbox{\hrule \hbox{\vrule height 0.2 cm
\hskip 0.2 cm \vrule height 0.2 cm}\hrule}\,}}
\newcommand{\dsl}{\pa \kern-0.5em /}

\def\slashchar#1{\setbox0=\hbox{$#1$}           
   \dimen0=\wd0                                 
   \setbox1=\hbox{/} \dimen1=\wd1               
   \ifdim\dimen0>\dimen1                        
      \rlap{\hbox to \dimen0{\hfil/\hfil}}      
      #1                                        
   \else                                        
      \rlap{\hbox to \dimen1{\hfil$#1$\hfil}}   
      /                                         
   \fi}

\begin{document}

\begin{titlepage}
\begin{center}

\hfill  DAMTP-2015-31

\vskip 1cm

{\Large \bf The Third Way to 3D Gravity}

\vskip 1cm

{\bf Eric Bergshoeff\,${}^1$, Wout Merbis\,${}^2$,}
\vskip .2truecm

{\bf Alasdair J. Routh\,${}^3$ and Paul K. Townsend\,${}^3$} \\

\vskip 25pt

{\em $^1$ \hskip -.1truecm
\em Centre for Theoretical Physics,
University of Groningen, \\ Nijenborgh 4, 9747 AG Groningen, The Netherlands\vskip 5pt }

{email: {\tt e.a.bergshoeff@rug.nl}} \\

\vskip .4truecm

{\em $^2$ \hskip -.1truecm  Institute for Theoretical Physics, Vienna University of Technology, \\
Wiedner Hauptstrasse 8-10/136, A-1040 Vienna, Austria\vskip 5pt }

{email: {\tt merbis@hep.itp.tuwien.ac.at}} \\

\vskip .4truecm

{\em $^3$ \hskip -.1truecm
\em  Department of Applied Mathematics and Theoretical Physics,\\ Centre for Mathematical Sciences, University of Cambridge,\\
Wilberforce Road, Cambridge, CB3 0WA, U.K.\vskip 5pt }

{email: {\tt A.J.Routh@damtp.cam.ac.uk; p.k.townsend@damtp.cam.ac.uk}} \\

\end{center}

\vskip 0.5cm

\begin{center} {\bf ABSTRACT}\\[3ex]
\end{center}
{ \small
Consistency of Einstein's gravitational field equation $G_{\mu\nu} \propto  T_{\mu\nu}$ imposes a ``conservation condition'' on the $T$-tensor that is satisfied by (i) matter stress
tensors, as a consequence of the matter equations of motion,  and (ii) identically by certain other tensors,  such as the metric tensor. However, there is a third way, overlooked
until now because it  implies  a ``non-geometrical'' action: one {\it not} constructed  from the metric and its derivatives alone. The new possibility  is exemplified by the 3D ``minimal massive gravity'' model,  which resolves the ``bulk vs boundary'' unitarity problem of  topologically massive gravity  with anti-de Sitter asymptotics.  Although all known examples of the third way are in three spacetime dimensions, the idea is general and could, in principle, apply to higher-dimensional  theories.
}
\vskip 1cm

\begin{center}
Essay written for the Gravity Research Foundation \\ 2015 Awards for Essays on Gravitation.

\end{center}

\end{titlepage}

\noindent
It is now 100 years since Einstein wrote down, after a long struggle,  his gravitational field equations, 
\begin{equation}\label{EFE}
G_{\mu\nu} = (8\pi G) \, T_{\mu\nu}\, ,
\end{equation}
where $T_{\mu\nu}$ is the matter stress tensor, $G$ is Newton's gravitational constant and $G_{\mu\nu}$ is  the Einstein tensor, defined in terms of the metric $g_{\mu\nu}$ and its Ricci tensor $R_{\mu\nu}$, and Ricci scalar $R$, by
\begin{equation}
G_{\mu\nu}= R_{\mu\nu} - \frac{1}{2} g_{\mu\nu} R\, .
\end{equation}
The  stress tensor satisfies, {\it as a consequence of the matter field equations}, the ``conservation condition''
\begin{equation}\label{conservation}
D^\mu T_{\mu\nu} =0\, ,
\end{equation}
where $D$ is the usual covariant derivative in a gravitational background. In  flat space this condition implies conservation of energy and momentum in the matter fields.
It was this condition that led Einstein to the Einstein tensor: he needed to construct from the metric a tensor $G_{\mu\nu}$  satisfying the Bianchi identity
\begin{equation}
D^\mu G_{\mu\nu} \equiv 0\, .
\end{equation}

However, one can turn the logic around: it is because the Einstein tensor satisfies the Bianchi identity
that the tensor $T_{\mu\nu}$ appearing on the right-hand side of the Einstein field equations (\ref{EFE}) must satisfy the conservation condition (\ref{conservation}).  This perspective leads to the following question. How many ways are there to construct  a tensor $T_{\mu\nu}$ satisfying
(\ref{conservation})?  Certainly, one may take $T_{\mu\nu}$ to be a matter stress tensor, but another obvious possibility  is to choose
$T_{\mu\nu} \propto  g_{\mu\nu}$.  The conservation condition is an {\it identity} for this tensor, which can be viewed as a dark-energy
contribution  to the  stress tensor, but it can also be taken over to the left-hand side to give us the modified field equation
\begin{equation}
G_{\mu\nu} + \Lambda g_{\mu\nu} = (8\pi G) T_{\mu\nu}\, ,
\end{equation}
where $\Lambda$ is a constant, the ``cosmological constant'' introduced by Einstein himself in 1917. But this is just a special case: for  any diffeomorphism
invariant functional $I[g]$,  the tensor
\begin{equation}
I_{\mu\nu} = \frac{1}{\sqrt{-\det g}} \frac{\delta I[g]}{\delta g^{\mu\nu}}
\end{equation}
satisfies the Bianchi-type identity
\begin{equation}
D^\mu I_{\mu\nu} \equiv 0\, .
\end{equation}
Taking all such tensors over to the left hand side
we arrive at a generalisation of the Einstein field equations of the form
\begin{equation}
\label{eequalst}
E_{\mu\nu} = (8\pi G) T_{\mu\nu}\, ,
\end{equation}
where $E_{\mu\nu}$ is a symmetric tensor satisfying $D^\mu E_{\mu\nu}\equiv 0$. It will have the form
\begin{equation}
E_{\mu\nu} = \Lambda g_{\mu\nu} + \sigma G_{\mu\nu} + \dots \, ,
\end{equation}
where $\sigma$ is a dimensionless number and the omitted terms are higher-dimension tensors with coefficients that have  dimensions of increasing powers of  inverse mass. 

To return to our question, we have now seen that there are two standard ways to construct a tensor satisfying the conservation condition (\ref{conservation}). Either:
\begin{itemize}
\item[1.] The condition is satisfied {\it as a consequence of the matter field equations}, and the tensor contributes to the matter stress tensor,
\item[] or:
\item[2.] It is satisfied {\it identically} by means of a Bianchi-type identity, and the tensor contributes to the gravitational tensor $E_{\mu\nu}$.
\end{itemize}
However, there is one other logical possibility; a ``third way''.  Any abstract discussion would make it appear very unlikely that it could be realised in practice, so we will instead present an example. The example is  ``minimal massive gravity'' (MMG) \cite{Bergshoeff:2014pca},  which  describes an interacting massive graviton in three spacetime dimensions (3D).

Our starting point will be the third-order field equation of ``topologically massive gravity'' (TMG) \cite{Deser:1981wh},  which propagates a single spin-2 mode of mass $\mu$. Allowing for a cosmological constant, the TMG field equation is \eqref{eequalst} with
\begin{equation}\label{TMGE}
E_{\mu\nu} = \Lambda g_{\mu\nu} + \sigma G_{\mu\nu} + \frac{1}{\mu} C_{\mu\nu}\, .
\end{equation}
Here, $C_{\mu\nu}$ is the Cotton tensor, defined as
\begin{equation}
C_{\mu\nu} = \frac{1}{\sqrt{-\det g}} \, \varepsilon_\mu{}^{\tau\rho} D_\tau S_{\rho\nu}\, ,
\end{equation}
where the $S$-tensor is the 3D Schouten tensor
\begin{equation}
S_{\mu\nu} = R_{\mu\nu} - \frac{1}{4}g_{\mu\nu} R\, .
\end{equation}
The Cotton tensor is symmetric and traceless, and it satisfies a Bianchi identity; the corresponding action is the Lorentz-Chern-Simons action for the  affine connection.

Much of the interest in 3D gravity models derives from the potential simplifications of a lower dimension for the quantum theory. For 3D models with anti-de Sitter (AdS) asymptotics  there is the prospect of defining a quantum gravity theory via a holographically dual 2D conformal field theory (CFT).  A necessary condition for unitarity in this context is that the (left/right) central charges of the CFT (which differ in a parity-violating theory like TMG) are both positive. Additional unitarity requirements are that the bulk  spin-2 mode be non-tachyonic and have positive energy.  Given a 3D gravitational  action, all these requirements can be checked in a semi-classical approximation:  by linearization about an AdS background  for the bulk spin-2 mode,  and by consideration of the asymptotic symmetry algebra \cite{Brown:1986nw} for the CFT central charges.  It turns out that there is no choice of TMG parameters for which all these unitarity conditions are satisfied \cite{Li:2008dq} (and the same is true of 
``new massive gravity'', which propagates a parity doublet of massive spin-2 modes \cite{Bergshoeff:2009hq}).  This is known as the ``bulk-boundary clash'' because the bulk mode has negative energy when both central charges of the boundary CFT are positive,  although all bulk unitarity conditions must also have a CFT interpretation. Attempts to circumvent this problem by including higher-derivative modifications of the $E$-tensor (\ref{TMGE}) do not succeed \cite{Sinha}\cite{Paulos}.

However, there is a third way to modify the TMG equation. To see this, consider the tensor
\begin{equation}
J_{\mu\nu} = \frac{1}{2\det g}\ \varepsilon_\mu{}^{\rho\sigma} \varepsilon_\nu{}^{\tau\eta} S_{\rho\tau}S_{\sigma\eta}\, .
\end{equation}
This does not satisfy the conservation condition identically; in fact
\begin{equation}
\sqrt{-\det g}\, D_\mu J^{\mu\nu} = \varepsilon^{\nu\rho\sigma} S_\rho{}^\tau C_{\sigma\tau} \not\equiv 0\, .
\end{equation}
However, if we use the source-free TMG equation to express the Cotton tensor as a linear combination of the Schouten tensor and the metric tensor, then we see that the right-hand-side of this equation vanishes identically. The $J$-tensor therefore satisfies the conservation condition {\it as a consequence of the TMG field equation}. This suggests the ``third way'':
\begin{itemize}
\item[3.] A tensor $T$ may satisfy  the conservation condition {\it as a consequence of the gravitational field equation itself}.
\end{itemize}
The obvious difficulty with this idea is that {\it we change the equation as soon as we include the new tensor as a source}. It seems that it would take a miracle for $J$ to continue to satisfy the conservation condition as a consequence of the gravitational field equation when $J$ itself is included  in this equation, but miracles of this type are not excluded!

For the case in hand, if we  take our source tensor to be
\begin{equation}
T_{\mu\nu} = - \frac{\gamma}{\mu^2} J_{\mu\nu}\, , 
\end{equation}
for some dimensionless constant $\gamma$, then the sourced equation implies that the Cotton tensor is now a linear combination of the Schouten tensor, metric tensor and the $J$-tensor. Using this expression, we find that
\begin{equation}
\sqrt{-\det g}\, D_\mu J^{\mu\nu} = -\frac{\gamma}{\mu}\varepsilon^{\nu\rho\sigma} S_\rho{}^\tau J_{\sigma\tau} \equiv 0 \,.
\end{equation}
The final identity is due to the specific  form of $J$ and the symmetry of the Schouten tensor, so the $J$-tensor is {\it still} conserved!  We have now verified the consistency of the ``minimal massive gravity'' equation
\begin{equation}\label{MMG2}
\Lambda g_{\mu\nu} + \sigma G_{\mu\nu} + \frac{1}{\mu} C_{\mu\nu} + \frac{\gamma}{\mu^2} J_{\mu\nu} =0\, , 
\end{equation}
which is still  a third-order  equation because the $J$-tensor is only second order.  This equation provides an explicit and simple example of the third way. It demonstrates that it may be possible to include,  in a gravitational field equation, tensors constructed entirely from the metric that do {\it not} satisfy a Bianchi identity.

In such cases there will be no ``geometrical'' action; i.e.~one constructed entirely from the metric. Nevertheless, at least for our MMG example, there {\it is} an action involving auxiliary fields \cite{Bergshoeff:2014pca}. We will not present it here, although it is fairly simple; instead we address the puzzle of how an action with auxiliary fields is possible when their elimination appears to give us an action for the metric alone, in contradiction to our claim that no such action exists. 

A set of  fields is ``auxiliary'' if the field equations allow us to solve for them  in terms of the other fields. The MMG action is such that the field equations determine the auxiliary fields in terms of the metric and its curvature tensor,  such that the remaining equation for the metric alone is precisely  (\ref{MMG2}). However,  back-substitution into the {\it action}  is legitimate only if the equations used to solve for the  auxiliary fields are equivalent, jointly, to the equations obtained by variation with respect to them. In the MMG case,  the equation obtained by variation of the metric is needed to solve for the  auxiliary fields, and this makes back-substitution into the action illegitimate. This will be the case for any action that yields   ``third-way consistent'' field equations.

To summarise: despite the fact that its left-hand side does {\it not} satisfy a Bianchi identity,  the MMG equation (\ref{MMG2}) {\it is} derivable from an action.  This could have been just a curiosity but it turns out to be essential to the resolution of the ``bulk-boundary clash'' of massive 3D gravity.  Linearization shows that MMG propagates a single massive spin-2 mode and a full Hamiltonian analysis shows that there are no other bulk degrees of freedom,  just like TMG. However, the AdS boundary properties of MMG differ from those of TMG: provided $\gamma<0$ one can choose parameters to arrange for both central charges of the asymptotic conformal symmetry algebra to be positive \cite{Bergshoeff:2014pca}.  Allowing for equivalences,  there is  a single connected region in the MMG parameter space for which all unitary conditions accessible to a semi-classical analysis are satisfied \cite{Arvanitakis:2014xna}.

Third-way consistency has another surprising consequence, which we again illustrate using MMG.  Let us add a source tensor ${\cal T}$ to the right-hand side of the MMG equation:
\begin{equation}
\Lambda g_{\mu\nu} + \sigma G_{\mu\nu} + \frac{1}{\mu} C_{\mu\nu} + \frac{\gamma}{\mu^2} J_{\mu\nu} =  \mathcal{T}_{\mu\nu} \,.
\end{equation}
For  TMG, we may choose ${\cal T}$ to be a matter stress tensor,  but this choice is not possible for non-zero $\gamma$; i.e.~for  MMG. In fact, consistency requires that
\begin{equation}\label{identity}
D_\mu {\cal T}^{\mu\nu} = \frac{\gamma}{\mu\sqrt{-\det g}}\,  \varepsilon^{\nu\rho\sigma} S_\rho{}^\tau {\cal T}_{\sigma\tau}\,  .
\end{equation}
Given a matter stress tensor $T$ one can find a corresponding source tensor ${\cal T}$ that solves this equation \cite{Arvanitakis:2014yja}, but it is quadratic  in $T$!
A corollary is that some solutions of MMG can exhibit qualitatively different behaviour when compared to analogous solutions of TMG because the
tensor ${\cal T}$ need not satisfy energy conditions imposed on  $T$. In particular, big-bang singularities of  TMG cosmological solutions  with an ideal fluid source are absent  in MMG \cite{Arvanitakis:2014yja}.

To conclude, we have illustrated the possibility of a ``third way'' to construct consistent gravitational field equations using the example of ``minimal massive gravity'', and we
have discussed its physical relevance in this context.  All known examples are in 3D, and there is an analog for 3D Yang-Mills theory \cite{Arvanitakis:2015oga},  but the basic idea is dimension independent. It may be  that there exist higher-dimensional gravitational models with field equations that are similarly third-way consistent. In view of this, it is perhaps worth recalling that one of the assumptions of Lovelock's generalisations  of the second-order Einstein field equations  to spacetimes of arbitrary dimension is that all terms satisfy a Bianchi identity \cite{Lovelock:1971yv}. The possibility of a ``third way'' in higher dimensions calls into question this assumption.

\section*{Acknowledgements}

We would like to thank Alex Arvanitakis and Olaf Hohm for their collaboration on work leading to this essay. AJR is supported by a grant from the London Mathematical Society, and he would also like to thank the University of Groningen for hospitality during the writing of this essay.

\providecommand{\href}[2]{#2}\begingroup\raggedright\endgroup

\end{document}